\documentclass[conference]{IEEEtran}
\IEEEoverridecommandlockouts
\usepackage{cite}
\usepackage{amsmath,amssymb,amsfonts}
\usepackage{algorithmic}
\usepackage{bbm}
\usepackage{graphicx}
\usepackage{algorithm}
\usepackage{algorithmic}
\usepackage{diagbox}
\usepackage{textcomp}
\DeclareMathOperator*{\argmax}{argmax} % thin space, limits underneath in displays
\usepackage{xcolor}
\usepackage{scalerel}
\usepackage{lettrine}
\usepackage{dsfont}
\usepackage{subcaption}
\usepackage{float}
\usepackage{amsmath,amsthm}
\usepackage{fixmath}
\usepackage[]{graphicx}
\usepackage{lipsum}
\def\BibTeX{{\rm B\kern-.05em{\sc i\kern-.025em b}\kern-.08em
		T\kern-.1667em\lower.7ex\hbox{E}\kern-.125emX}}

\begin{document}
\title{Multi-Agent Deep Stochastic Policy Gradient for Event Based Dynamic Spectrum Access\\
\thanks{Rahif Kassab completed this work while an intern at Huawei's Mathematical and Algorithmic Sciences Lab.}}

\author{\IEEEauthorblockN{Rahif Kassab\IEEEauthorrefmark{2},
Apostolos Destounis\IEEEauthorrefmark{1}, Dimitrios Tsilimantos\IEEEauthorrefmark{1} and Mérouane Debbah\IEEEauthorrefmark{1}\\}
\IEEEauthorblockA{\IEEEauthorrefmark{2}King's Communications, Learning \& Information Processing (KCLIP) Lab, King's College London,
London, UK}
\IEEEauthorblockA{\IEEEauthorrefmark{1}\small Mathematical and Algorithmic Sciences Lab, Paris Research Center, Huawei Technologies Co. Ltd.\\ Emails:\IEEEauthorrefmark{2}\{firstname.lastname\}@kcl.ac.uk, \IEEEauthorrefmark{1}\{firstname.lastname\}@huawei.com }\\}

\maketitle

\begin{abstract}
We consider the dynamic spectrum access (DSA) problem where $K$ Internet of Things (IoT) devices compete for $T$ time slots constituting a frame. Devices collectively monitor $M$ events where each event could be monitored by multiple IoT devices. Each device, when at least one of its monitored events is active, picks an event and a time slot to transmit the corresponding active event information. In the case where multiple devices select the same time slot, a collision occurs and all transmitted packets are discarded. In order to capture the fact that devices observing the same event may transmit redundant information, we consider the maximization of the average sum event rate of the system instead of the classical frame throughput. We propose a multi-agent reinforcement learning approach based on a stochastic version of Multi-Agent Deep Deterministic Policy Gradient (MADDPG) to access the frame by exploiting device-level correlation and time correlation of events. Through numerical simulations, we show that the proposed approach is able to efficiently exploit the aforementioned correlations and outperforms benchmark solutions such as standard multiple access protocols and the widely used Independent Deep Q-Network (IDQN) algorithm.
\end{abstract}
\begin{IEEEkeywords}
	Multi-Agent Reinforcement Learning, Dynamic Spectrum Access, Throughput Maximization
\end{IEEEkeywords}
%%%%%%%%%%%%%%%%%%%%%%%%%%%%%%%%%%%%%%%%%%%%%%%%%%%%%%%%%%%%%%%%%%%
\IEEEpeerreviewmaketitle
\section{Introduction}
\label{sec:introduction}
Current static spectrum allocation techniques, such as Time Division Multiple Access (TDMA) and Orthogonal Frequency Division Multiple Access (OFDMA) %\cite{OFDM_comparison_LTE_WiMAX} 
fall short when the number of connected devices increases. More specifically, these solutions lead to an inefficient use of radio resources when allocated resources are not utilized due to possible idle states of corresponding devices. This problem is aggravated with 5G and beyond technologies, namely massive Machine Type Communications (mMTC) \cite{mmtc_5g_physical_solutions}, which are characterized by their sporadic traffic. Accordingly, allocating dedicated radio resources for each device may no longer be an efficient solution. Hence, in recent years,
Dynamic Spectrum Access (DSA) has been investigated both in academia and industry as a more flexible and promising alternative to static spectrum allocation schemes \cite{DSA_survey}. In DSA, devices can choose any time slot in a given frame to transmit their packet. In this way, radio resources are used more efficiently as they are not wasted by being allocated to devices that could be idle. However, one major drawback is that collisions can occur when multiple devices choose to transmit in the same radio resource. \par Consequently, extensive research was done in past years to derive protocols that maximize device throughput, the most famous being slotted ALOHA \cite{aloha_1970}. Since ALOHA was proposed in $1970$, extensive research was conducted to improve its performance. For brevity, we only mention here the most recent major variation to ALOHA, namely the coded slotted ALOHA \cite{coded_slotted_aloha} based on linear block encoding and Successive Interference Cancellation (SIC). However, this technique comes with multiple caveats like hardware and software changes and higher energy consumption at both the transmitter and the receiver to account for coding and SIC. Furthermore, this technique cannot adapt to dynamic environments, e.g., variable number of devices and channel characteristics.\par
The problem of DSA has been investigated from multiple perspectives including Bandits \cite{anima_distributed_algos_cognitive}, game theory \cite{walid_gametheory_wireless} and matching theory \cite{matching_theory_walid}. All these studies, however, derive transmission strategies that only work for static environments, i.e., fixed number of devices and channel switching patterns, for a certain utility (mostly, individual throughput) and more importantly require prior information about the environment and the communication system (for e.g., environment transition matrix). We refer to \cite{DSA_survey} for a more comprehensive survey.\par
Recently, Reinforcement Learning (RL) \cite{sutton_book} has been studied as a potential efficient solution for the DSA problem. The motivation for using RL is mainly to efficiently adapt to environment dynamics and changing activation patterns. Accordingly, a deep Q-Network (DQN) is considered in \cite{DRL_DMA_trans_cog} to find a transmission strategy, commonly referred to as policy, that maximizes the long-term number of successful transmissions. However, it considers a single user and a simplistic channel model where the latter could be in a "good" or "bad" state. Furthermore, DSA has been studied in \cite{yu_DRL_MA_JSAC} in the context of heterogeneous networks by using a DQN for throughput maximization. However, a gateway is needed to take decisions after collecting rewards from all agents which increases the communication overhead. Furthermore, authors in \cite{kobi_twc} consider a multi-user scenario and use DQN augmented with a Long Short-Term Memory. %to learn from the agent's history, which is very effective in Partially Observed Markov Decision Processes (POMDP) scenario \cite{deep_recurrent_q_learning}. 
Similar works using Q-learning and DQN can be found respectively in \cite{RLMAC_elhanany} and \cite{buehrer_deep}. Finally, an Actor-Critic framework using deep neural network (DNN) is analyzed in \cite{zhong_AC_DMA}. \par
In contrast to all previous work, we consider here a more realistic scenario where devices are monitoring events and each event can be active or not according to a Markov chain with \textit{unknown} transition probabilities. Accordingly, device activities can be correlated in time. Furthermore, as multiple devices can monitor the same event, device activities are also spatially correlated (named here device-level correlation). 
%In contrast to most previous works, and inline with \cite{zhong_AC_DMA}, we use policy gradient RL and more specifically Actor-Critic due to its scalability properties over value-based RL techniques like DQN that focuses solely on estimating the value function. Furthermore, in contrast to previous work on DSA, we consider here correlated devices activity where devices observing the same underlying event could be active at the same time. 
A similar model was considered in \cite{common_alarm_petar} and \cite{rahif_grantfree} where the goal was to correctly detect underlying phenomena observed by devices. To the best of the authors' knowledge, this is the first paper using RL to exploit temporal and spatial correlation of underlying device observations for DSA.
% For extension to COMA
%However, classical actor-critic algorithms \cite{AC_algorithms} are not suitable for multi-user scenarios. The main reason is that training is done independently at each device, i.e., without considering other devices actions, which makes it difficult to learn coordinated transmission strategies. Furthermore, it suffers from the credit assignment problem \cite{credit_assignment} which is more critical in DSA problems. Consequently, we propose in this paper the use of Counter Factual Multi-Agent Policy Gradients (COMA) \cite{coma} as a potential attractive solution to the DSA problem, hence the name of our proposed algorithm COMA-DSA. The main intuitions behind COMA are two folds, first, as training is done offline, it makes more sense to use a centralized ciritc that conditions over the actions-observations history of all the agents. Second, when optimizing a common reward (in our case, sum throughput), it is not possible for each agent with the current actor-critic algorithms to learn the contribution of its action in the global reward. Of course, individual rewards (for e.g., individual rates) can be used, however, in this case, each agent could end up acting greedily and it won't be possible to learn cooperative strategies that will benefit all the agents. \\
\par The rest of the paper is organized as follows: in Sec. \ref{sec:system_model_and_metrics} we detail the system model and the performance metrics. In Sec. \ref{sec:reinforcement_learning} we formulate the RL problem and detail two RL algorithms. Benchmark protocols and experiments are discussed in Sec. \ref{sec:numerical_results}. Finally, conclusions are drawn in Sec. \ref{sec:conclusions}.
%%%%%%%%%%%%%%%%%%%%%%%%%%%%%%%%%%%%%%%%%%%%
\begin{figure*}[t]
	\centering
	\includegraphics[height= 9 cm, width= 17 cm]{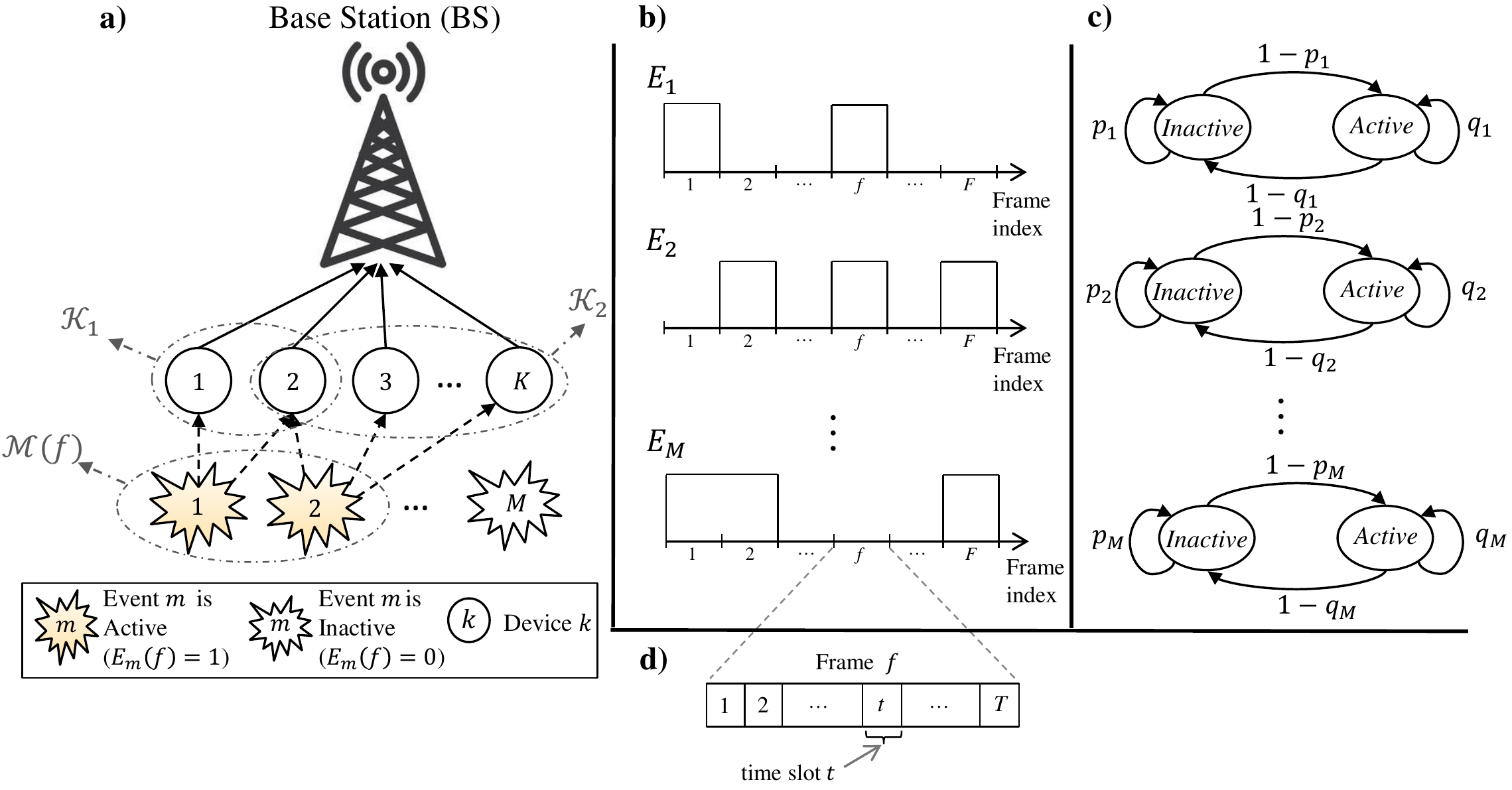}
	\caption{a) IoT system with $K$ devices, each monitoring a subset of $M$ events and reporting to a BS, b) Each event $m$ is either active or inactive in each one of the frames, c) Dynamic behavior of event activation is defined by independent Markov chains with unknown transition probabilities, d) Frame with $T$ time slots where a device picks a time slot to transmit information about an active event using an optimized policy.}
	\label{fig:system_model}
\end{figure*}
%%%%%%%%%%%%%%%%%%%%%%%%%%%%%%%%%%%%%%%%%%%
\section{System Model and Performance metrics}
\label{sec:system_model_and_metrics}
\subsection{System Model}
\label{sec:system_model}
As illustrated in Fig. \ref{fig:system_model}(a), we consider a radio access scenario in which $M$ events are monitored by $K$ devices, such as IoT devices or wireless sensors, which are connected to a Base Station (BS). An event can be, for instance, the detection of a high temperature level. Each device $k \in \{1,2,\ldots,K \}$ can monitor multiple events, but can transmit corresponding information about only one active event at any given time. The state of the $m$-th event in frame $f=1,2,\ldots,F$ is defined by a random variable $E_m(f)$ for $m=1,\ldots,M$, which determines whether the event is active ($E_m(f)=1$) or inactive ($E_m(f)=0$). The activation pattern of $E_m(f)$, for $f=1,\ldots, F$ evolves according to a Markov chain with \textit{unknown} switching probabilities $p_m$ and $q_m$ as shown in Fig. \ref{fig:system_model}(b)-(c). The event variables $E_m (f)$ are independent across different events. More general Markov chains and correlation assumptions can be directly accommodated for in the framework. Each event $m$ is monitored by a subset $\mathcal{K}_m \subseteq \{1,2,\ldots,K\}$ of devices, with each device $k$ monitoring a subset $\mathcal{M}_k \subseteq \{1,2,\ldots,M\}$ of events. Each device is unaware of the subset of events monitored by other devices.
\par
As shown in Fig. \ref{fig:system_model}(b) and (d), time is divided into frames, indexed as $f \in \{1, \ldots, F \}$, each composed of $T$ time slots, which are indexed in turn as $t \in \{1,\ldots, T \}$. At the beginning of each frame $f$, a subset $\mathcal{M}(f) \subseteq \{1,2,\ldots,M\}$ of events is active, i.e., $\mathcal{M}(f)= \{m :\ E_m(f)= 1 \}$. 
%Each device $k$ observing any of the active events can transmit an event with some unknown probability $p_k$, e.g., due to insufficient energy or malfunctioning.
Each transmitting device picks a time slot $t$ among the $T$ time slots in frame $f$ to deliver one packet. When a device observes multiple active events, its transmitted packet contains information about only one of them due to payload size limitations. \par
When multiple devices transmit in the same time slot, a collision occurs and thus the BS cannot recover any of the packets sent. Moreover, if multiple devices transmitted successfully information about the same active event within a given frame, the event is considered redundant and thus counted only once at the BS in terms of the adopted performance metric. At the end of each frame, each device gets an acknowledgment (ACK) message indicating whether its own transmission was successful, redundant or collided with other device transmissions. The goal of this work is to devise learning strategies that enable active devices to optimize, in a decentralized fashion, the policy that selects time slot $t$ and the event to be communicated based on the history of received ACK signals. It is noted that a successful strategy should implicitly learn and exploit time and spatial correlations among the activation of devices, which in turn depend on the unknown subsets of monitored events by other devices, event statistics and transmission probabilities based solely on the ACK signal.\par
For each transmitting device $k$, we denote as $x_{k,m,t}(f)$ the indicator variable that is equal to $1$ when device $k$ transmits information about event $m$ in time slot $t$ of frame $f$. Note that we can have $x_{k,m,t}(f)=1$ only if device $k$ belongs to set $\mathcal{K}_m$ for an active event $m \in \mathcal{M}(f)$, and $0$ otherwise.
%%%%%%%%%%%%%%%%%%%%%%%%%%%%%%%%%%%%%%%%
\subsection{Performance Metrics}
\label{sec:performance_metrics}
We define the system's \textit{event rate} for each frame $f$ as the fraction of active events $\mathcal{M}(f)$ whose information is correctly received at the BS in the $f$-th frame. Mathematically, this can be expressed as
\begin{equation}
	R(f) = \frac{\sum_{m \in \mathcal{M}(f)} \mathbbm{1}_{\{ c_m(f) \geq 1\}}} {\min (T,|\mathcal{M}(f)|)}, \label{eq:R_f}
\end{equation}
where the normalization is over the maximum number of events that can be correctly received in a frame. $\mathbbm{1}_{\{.\}}$ is the indicator function and $c_m(f)= \sum_{t=1}^{T} c_{m,t}(f)$ counts the number of time slots in which event $m$ is successfully reported, i.e., 
\begin{align}
    	c_{m, t}(f) = 1\ \mathrm{if}\ \sum_{k \in \mathcal{K}_m} x_{k,m,t}(f) = 1\ \mathrm{and}\ \label{eq:c_m}\\ \sum_{\substack{m^\prime \in \mathcal{M}(f)\\ m^\prime \neq m }} \sum_{k \in \mathcal{K}_{m^\prime} } x_{k,m^{\prime},t}(f) = 0. \nonumber 
\end{align}
In words, $c_{m,t}(f)=1$ if a single device transmits in slot $t$ as long as such device transmits information about event $m$. Note that, by using \eqref{eq:R_f} and \eqref{eq:c_m} redundant packets containing information about the same event are counted only once. The average event sum rate can then be written by summing over all frames as 
\begin{equation}
	R = \frac{\sum_{f=1}^{F}R(f)}{F}.
\end{equation}
The joint goal of all devices is to maximize the expected discounted average sum rate, i.e., 
\begin{equation}
\underset{\{\pi_k\}_{k=1}^{K}}{\mathrm{maximize}}\ \mathbb{E}\Big[\sum_{f \geq 1} \gamma^{f} R(f)\Big], \label{eq:optimization}
\end{equation}
where the expectation is taken over the distribution of activities of events and devices. In \eqref{eq:optimization}, the factor $\gamma \in (0,1)$ discounts the current value of future rates \cite{sutton_book}. The maximization is taken over the policies $\{\pi_k\}_{k=1}^{K}$, where the policy $\pi_k$ of each device $k$ maps its observation of ACK messages and previous actions $\{x_{k,m,t} (f^\prime)\}_{\substack{m \in \mathcal{M}_k \\ t \in \{1, \ldots, T\}}}$ for $f^\prime < f$ to a probability distribution over the current actions $\{x_{k,m,t} (f)\}_{\substack{m \in \mathcal{M}_k \\ t \in \{1, \ldots, T\}}}$. Since the objective function \eqref{eq:optimization} depends on the actions of all other devices from the viewpoint of each device, the problem belongs to the family of Decentralized Partially Observable Markov Decision Process (Dec-POMDP) \cite{review_MARL}. \\
%%%%%%%%%%%%%%%%%%%%%%%%%%%%%%%%%%%%%%%%%%%%%%%%%%%%%%%%%%%%%%%%%%%%%%%%%%
\section{Reinforcement Learning Solution}
\label{sec:reinforcement_learning}
\subsection{Formulation}
Due to its complexity, we tackle \eqref{eq:optimization} with the framework of Multi-Agent Reinforcement Learning (MARL) by leveraging neural networks as function approximators.\par
Specifically, at the beginning of each frame $f$, each device $k$ observes the $|\mathcal{M}_k|\times 1$ binary vector $\mathbf{o}_k(f)$ defined as follows
\begin{equation}
\mathbf{o}_k (f) = \{E_m(f) \}_{m \in \mathcal{M}_k}
\end{equation}
indicating which monitored events are active. Then, each transmitting device $k$ decides which active monitored event $m_k(f) \in \mathcal{M}_k (f) = \mathcal{M}_k \cap \mathcal{M}(f)$ and in which slot $t_k(f) \in \{0,1,\ldots,T\}$ to transmit, where $0$ denotes the case where the device prefers to stay idle (this is useful to avoid collisions with other devices transmissions). Accordingly, the action of device $k$ is defined by the tuple $\mathbf{a}_k(f)= (m_k(f), t_k(f)) \in \mathcal{A}_k(f) = \mathcal{M}_k(f) \times \{0,1,\ldots,T \} $. Note that we have $x_{k,m,t}(f)=1$ for $m=m_k(f)$ and $t=t_k(f)$ and $0$ otherwise. Each device $k$ also keeps track of its previous actions and observations. More precisely, the history of observations and actions of device $k$ at frame $f$ is given by  $\mathbf{h}_k(f) = (\mathbf{o}_k(f-1),\mathbf{a}_k(f-1),\ldots,\mathbf{o}_k(f-W), \mathbf{a}_k(f-W))$, where a sliding window of size $W$ is applied in order to keep memory requirements limited to the $W$ most recent frames. Alternatively, one could let $W$ unconstrained by leveraging recurrent neural networks \cite{deep_recurrent_q_learning}. The device's policy $\pi_k$ maps history $\mathbf{h}_k(f)$ into action $\mathbf{a}_k(f)$ using a DNN approximator, as detailed below. In order to drive learning at each device towards a solution that approximately optimizes problem \eqref{eq:optimization}, we adopt a reward signal $r_k(f)$ for each device $k$ such that: \\ \\
$r_k(f)$ = $
\begin{cases}
	A\ \text{if device $k$'s transmission is successful and}\\ \text{the transmitted event is not redundant} \\
	B\ \text{if device $k$'s transmission is successful and}\\ \text{the transmitted event is redundant}\\
	C\ \text{if a collision occurs} \\
	0\ \text{otherwise},
\end{cases}$
where a redundant event means that information about the given event was transmitted successfully by another device. In general, we have $C < B 
\leq 0$ in order to discourage actions leading to collisions or redundant transmissions, while $A> 0$.
%%%%%%%%%%%%%%%%%%%%%%%%%%%%%%%%%%%%%%%%
% \setlength{\textfloatsep}{0.5cm}
% \setlength{\floatsep}{0.1cm}
\begin{algorithm}[t]
	\caption{Independent DQN (IDQN) Algorithm for each device $k \in \{1,2,\ldots, K\} $}
	\begin{algorithmic}[1]
		\renewcommand{\algorithmicrequire}{\textbf{Input:} History window size $W$, number of slots $T$, $\gamma, \epsilon_{min}$. }
		\renewcommand{\algorithmicensure}{\textbf{Output:}}
% 		\REQUIRE 
		\ENSURE  Device $k$'s policy $\pi_k: \mathbf{h}_k \rightarrow \mathbf{a}_k$.
	%	\\ \textit{Initialization: }$\mathbold{\theta}_k(1)\sim \mathcal{N}(0,0.02)$, $\gamma = 0.95$, $\epsilon=0.2$, $d=1.001$, $\mathcal{H}_k(f)=()$
		\\ \textit{Initialization: }$\mathbold{\theta}_k(1)$, $\epsilon$, $h_k(0)=\emptyset$ 
		\FOR{$f=1$ to $W$}
		\STATE get observation $\mathbold{o}_k(f)$
		\STATE take a uniformly distributed random action $\mathbf{a}_k(f) \in \mathcal{A}_k(f)$
		\STATE update History $\mathbf{h}_k(f) \leftarrow (\mathbf{h}_k(f-1), \mathbold{o}_k(f),\mathbf{a}_k(f))$
		\ENDFOR
		\REPEAT
		\STATE $f := f + 1 $
		\STATE get observation $\mathbf{o}_k(f)$
		\STATE sample $z \sim\ \mathrm{Bernoulli}(\epsilon)$
		\IF {$z == 1$ (exploration)}
		\STATE select action $\mathbf{a}_k (f) \in \mathcal{A}_k(f)$ uniformly at random
		\ELSE
		\STATE select $\mathbf{a}_k(f) = \argmax_{\mathbf{a} \in \mathcal{A}_k(f)} Q(\mathbf{h}_k(f) ,\mathbf{a}; \mathbold{\theta}_k(f))$
		\ENDIF
		\STATE update exploration rate: $\epsilon \leftarrow \max (\epsilon \times \epsilon, \epsilon_{min})$
		\STATE get reward $r_k(f)$
		\STATE update History: $\mathbf{h}_k(f+1) \xleftarrow{} (\mathbf{h}_k (f), \mathbf{o}_k(f), \mathbf{a}_k(f))$
		\STATE set target $y_k := r_k(f) + $ \\ \ \ \ \ \ \ \ \ \ \ \ $ \gamma \max_{\mathbf{a} \in \mathcal{A}_k(f)} Q(\mathbf{h}_k(f+1),\mathbf{a};\mathbold{\theta}_k(f))$
		\STATE update DQN weights $\mathbold{\theta}_k(f+1)$ by performing SGD on $(y_k - Q(\mathbf{h}_k(f), \mathbf{a}; \mathbold{\theta}_k(f)))^2$ 
		\UNTIL{convergence}
		\RETURN $\mathbold{\theta}_k^{\star}$
	\end{algorithmic} \label{algo_DQN}
\end{algorithm}
%%%%%%%%%%%%%%%%%%%%%%%%%%%%%%%%%%%%%%%%%%%%%%%%%%%%%%%%$$$
\subsection{Independent DQN Algorithm}
Based on the formulation in the previous section, we use as a benchmark the $\epsilon$-greedy independent DQN algorithm \cite{review_MARL} detailed in Algorithm \ref{algo_DQN} given its wide usage in the literature. The latter is executed independently at each device by considering the actions of the other devices as part of the environment as in \cite{two_agent_DQN}. The algorithm relies on training a DNN with weights $\mathbold{\theta}_k$ that aims to approximate a function, named $Q$-function, by minimizing its mean squared distance with a target value. The Q-function models the goodness and badness of taking a certain action given the history of past observations and actions.\\
More specifically, each device starts by taking $W$ random actions in the first $W$ frames and getting corresponding rewards in order to fill the history buffer (\textbf{lines 1-5}). Then, for each new frame (i.e., new observation), the $\epsilon$-greedy algorithm is used to take an action uniformly at random with probability $\epsilon < 1$ (\textbf{line 11}) or a greedy action, i.e. the action with the highest $Q$-function value (obtained at the output of the DNN), with probability $1-\epsilon$ (\textbf{line 13}). $\epsilon$ usually decreases slowly from one episode to another (\textbf{line 15}). We set the minimum value of $\epsilon$ to $\epsilon_{min}$ in order to ensure that each device visits all possible state-action pairs for a sufficiently high number of episodes. Then, after getting the corresponding reward and updating the history (\textbf{lines 16-17}), the DNN parameter $\mathbold{\theta}_k$ is updated using stochastic gradient descent (SGD) to minimize the mean squared error between the $Q$-function value of the executed action and a target action (\textbf{lines 18-19}).
\begin{algorithm}[t]
	\caption{Multi-Agent Deep Stochastic Policy Gradient (MADSPG) Algorithm}
	\begin{algorithmic}[1]
		\renewcommand{\algorithmicrequire}{\textbf{Input:} number of slots $T$, $\gamma$, target networks update rate $\tau$ }
		\renewcommand{\algorithmicensure}{\textbf{Output:}}
		\REQUIRE 
		\ENSURE  Device $k$'s policy $\pi_k: \mathcal{A}_k \rightarrow [0;1]$.
		\\ \textit{Initialization: }actor and target actor parameters $\mathbold{\theta}_{k}^{\pi}(1)=\mathbold{\theta}_{k}^{\pi^\prime}(1)$, critic and target critic parameters $\mathbold{\theta}_{k}^{\mu}(1)=\mathbold{\theta}_{k}^{\mu^\prime}(1)$
		%	\\ \textit{Initialization: }$\mathbold{\theta}_k(1)\sim \mathcal{N}(0,0.02)$, $\gamma = 0.95$, $\epsilon=0.2$, $d=1.001$, $\mathcal{H}_k(f)=()$
		\FOR{$f=1$ to $F$}
		\STATE \textbf{for} $k \in \{1, \ldots,K \}$: get observation $\mathbold{o}_k(f)$, take action $\mathbf{a}_k(f) \sim \pi_{k}$, get reward $r_k(f)$ and get new observations $\mathbold{o}^{\prime}_{k}(f)$, store $\{\mathbold{o}_k(f), \mathbf{a}_k(f),  r_k(f),\mathbold{o}_{k}^{\prime}(f)\}_{k=1}^{K}$ in replay buffer 
		\REPEAT
		\FOR{$k=1$ to $K$}
		\STATE sample a random minibatch $\{\mathbold{o}_i^s(f), \mathbf{a}_i^s(f),  r_i^s(f),\mathbold{o}_{i}^{\prime s}(f)\}_{s=1,i=1}^{S,K}$ of $S$ samples from replay buffer
		\STATE set critic target: $y^s_k =r^s_k + \gamma Q_k^{\mu^\prime}(\{\mathbold{o}_{i}^{\prime s}(f), \mathbf{a^\prime}_i^s(f) \}_{i=1}^{K}) \Big|_{\mathbf{a}_i^s(f) \sim \pi^\prime_i(\mathbold{o}_{i}^{ s}(f))}  $
		\STATE update critic DNN weights $\theta_{k}^{\mu}$ by minimizing the loss $L (\mathbold{\theta}^{\mu}_k) =$ \\$ \frac{1}{S} \sum_{s=1}^{S} \Big(y_k^s - Q_k^{\mu}( \{ \mathbold{o}_i^s(f), \mathbf{a}_i^s(f)  \}_{i=1}^{K})\Big)^2$
		\STATE update actor DNN weights $\theta_k^{\pi}$ using policy gradient: 
		 $\nabla_{\mathbold{\theta}_k^{\pi}}J(\mathbold{\theta}_k^{\pi}) \approx \frac{1}{S}\sum_{s=1}^{S} \nabla_{\mathbold{\theta}_k^{\pi}} \log \pi_k (\mathbf{o}_k^s(f)) Q^\mu_k(\{\mathbold{o}_k^s,\mathbold{a}_k^s \}_{k=1}^{K})\ \text{for}\ $\\ $\mathbf{a}_k^s \sim \pi_k (\mathbf{o}_k^s(f)) $
		\ENDFOR
		\STATE \textbf{for} $k \in \{1, \ldots, K\}$: update target critic and target actor : $\mathbold{\theta}_{k}^{\pi^\prime}(f+1) \leftarrow \tau\mathbold{\theta}_{k}^{\pi^\prime}(f) + (1-\tau)\mathbold{\theta}_{k}^{\pi}(f)$, $\mathbold{\theta}_{k}^{\mu^\prime}(f+1) \leftarrow \tau\mathbold{\theta}_{k}^{\mu^\prime}(f) + (1-\tau)\mathbold{\theta}_{k}^{\mu}(f)$
		\UNTIL{convergence}
		\ENDFOR
		\RETURN $\mathbold{\theta}_k^{\pi \star}$
	\end{algorithmic} \label{algo_MADSPG}
\end{algorithm}
%%%%%%%%%%%%%%%%%%%%%%%%%%%%%%%%%%%%%%%%%%%%%%
\subsection{MADSPG Algorithm}
\label{sec:madspg_algorithm}
IDQN has good performance in general in single agent scenarios as it relies on the theoretical convergence guarantees of the $Q$-learning algorithm in Markov Decision Process problems \cite{sutton_book}. However, as our problem belongs to the family of Dec-POMDP, we propose MADSPG in Algorithm \ref{algo_MADSPG} which is based on the recent advances in MARL, particularly the Multi-Agent Deep Deterministic Policy Gradient (MADDPG) algorithm introduced in \cite{MADDPG}. \\
\textit{Comparison with IDQN:} MADSPG belongs to the actor-critic family of RL algorithms. Instead of relying solely on the $Q$-function to evaluate the goodness of actions and also selecting next actions as in IDQN, MADSPG uses the following distinct components:
\begin{itemize}
    \item Actor: modelled by a DNN with weights $\mathbold{\theta}_{k}^{\pi}$. Its role is to approximate a stochastic policy $\pi_k(\mathbf{o}_k(f))$ that maps the observations of each device in each frame to a probability over the action space $\mathcal{A}_k$.
    \item Critic: modelled by a DNN with weights $\mathbold{\theta}_{k}^{\mu}$. Its role is to approximate the $Q$-function (as in IDQN). It is trained in a centralized manner, i.e., assumes that each device has information about all other devices actions and observations, while execution is done in a decentralized fashion. This can be made possible by having an exchange mechanism between devices during training or via signaling with the BS.
    \item Target Actor and target critic: modelled by two distinct DNNs with weights $\mathbold{\theta}_{k}^{\pi^{\prime}}$ and $\mathbold{\theta}_{k}^{\mu^{\prime}}$ respectively. The role of these networks is to improve stability during learning. More information can be found in \cite{MADDPG}.
\end{itemize}
\textit{Comparison with MADDPG: } MADSPG uses a discrete action space instead of a continuous one, in addition to a stochastic gradient update which makes it more suitable to learn stochastic policies given the randomness of each device observations.\par
More concretely, the algorithm starts by filling up a replay buffer (\textbf{line 2}) by taking an action according to the current policy $\pi$. Then, it samples a minibatch of $S$ samples from the buffer (\textbf{line 5}). The target critic value is then computed and the critic weights $\mathbold{\theta}_{k}^{\mu}$ are updated accordingly (\textbf{lines 6-7}). The actor is then updated using the policy gradient theorem (\textbf{line 8}) \cite[Section 13.2]{sutton_book} and finally the target networks are updated via soft updates using the update parameter $\tau$ (\textbf{line 10}).
%%%%%%%%%%%%%%%%%%%%%%%%%%%%%%%%%%%%%%%%%
%\section{COMA-DSA algorithm}
\section{Experiments}
\label{sec:numerical_results}
In this section, we provide numerical experiments in which we compare the performance of IDQN and the proposed MADSPG algorithm with the following benchmark transmission strategies:
\begin{itemize}
	\item TDMA: in each frame of $T$ time-slots, $T$ random devices are scheduled. When multiple events monitored by a given device are active, a random event is selected for transmission in the allocated time-slot;
	\item ALOHA: each device reports a random active event at a randomly selected time-slot.
\end{itemize}
% We note that, while used as benchmarks in other works \cite{DRL_DMA_trans_cog}\cite{zhong_AC_DMA}, Myopic \cite{myopic_zhao} and Whittle index \cite{whittle_index} based heuristic transmission strategies are not applicable for our setup. Indeed, both policies assume channels with i.i.d. switching patterns and prior knowledge of system dynamics. \par
In order to study the effect of time correlation on the system's performance, we plot the average events sum rate \eqref{eq:optimization} as function of $1-\mu$, where $\mu = 1-p-q$ is the time correlation coefficient \cite{correlation_markov} with $p \leq 1/2$ and $q \leq 1/2$ where $p_m=p$ and $q_m = q$ for all events $m \in \mathcal{M}$. In fact, a higher value of $\mu$ implies a higher time correlation between active and inactive states for each event. Indeed, $\mu = 1$ corresponds to the full correlation case, i.e., a deterministic switching where each active state is followed by an inactive state and vice versa while $\mu=0$ corresponds to the absence of time correlation.
\begin{figure*}[t!]
	\centering
	\begin{subfigure}[t]{0.5\textwidth}
	\centering
	\includegraphics[height= 5 cm, width= 4 cm]{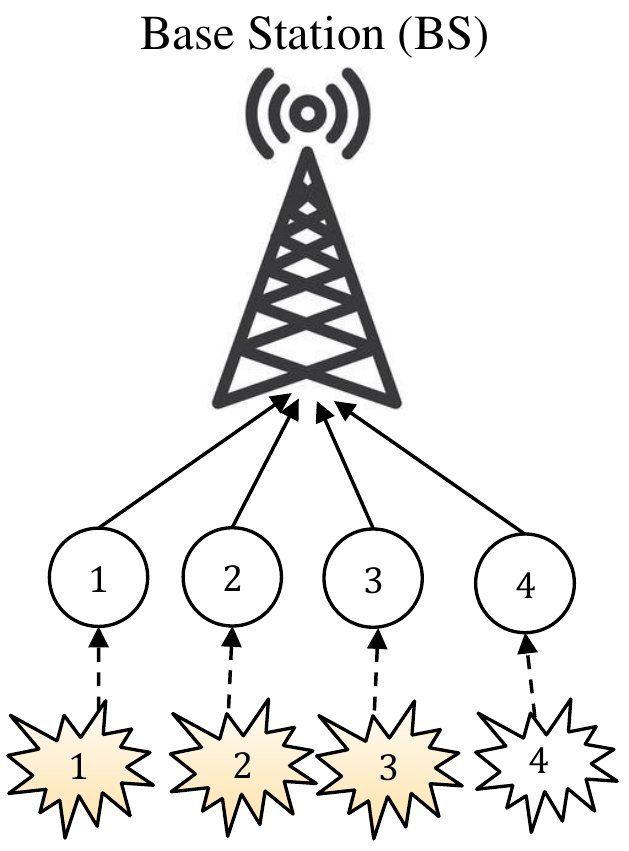}
	\caption{No device-level correlation.}
	\label{fig:toy_example_no_correlation}
	\end{subfigure}%
	~ 
	\begin{subfigure}[t]{0.5\textwidth}
	\centering
	\includegraphics[height= 5 cm, width= 4 cm]{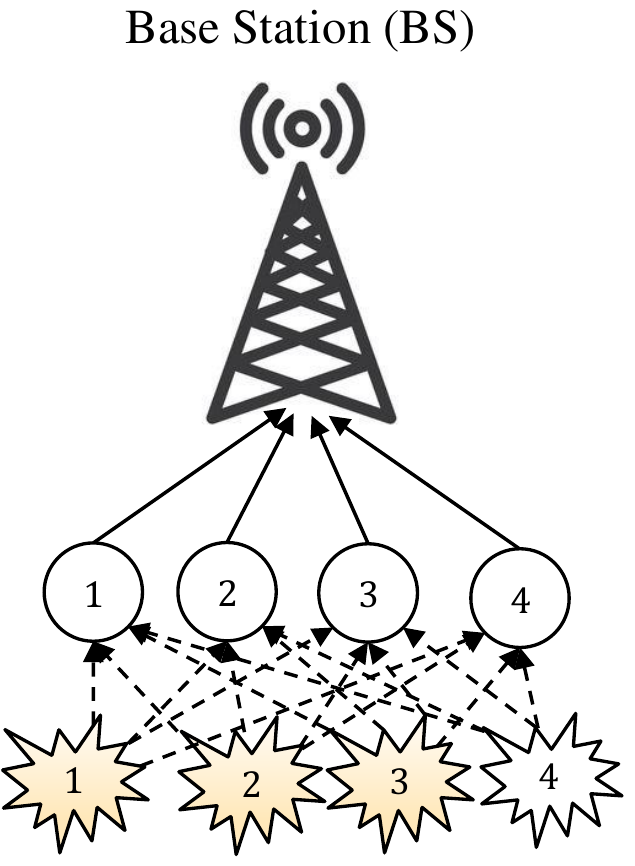}
	\caption{Full device-level correlation.}
	\label{fig:toy_example_full_correlation}
	\end{subfigure}
	\caption{Scenarios considered in Sec. \ref{sec:numerical_results} with $M=4$ Markovian events monitored by $K=4$ devices using frames of $T=2$ time slots with (a) no device-level correlation and (b) full device-level correlation.}
	\vspace{-5 mm}
\end{figure*}
\begin{figure}[t]
	\centering
	\includegraphics[height= 6 cm, width= 7.5 cm]{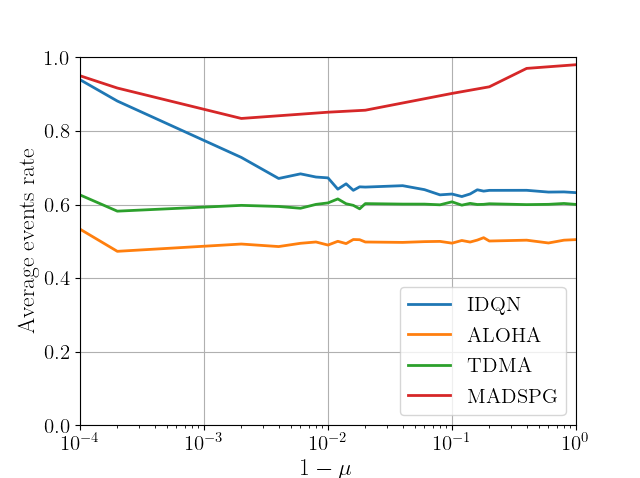}
	\caption{Average events rate for no device-level correlation scenario.}
	\label{fig:events_rate_no_space_correlation}
	\vspace{- 5mm}
\end{figure}
Numerical values are set as follows: $\epsilon = 0.9,\ \epsilon_{min}=0.05,\ \gamma=0.9,\ K=4,\ M=2,\ T=2,\ A=10,\ B=-5,\ C=-10$. We run each algorithm for $2000$ episodes averaged over $50$ runs. \par We first consider the set-up in Fig. \ref{fig:toy_example_no_correlation} where each device monitors only one event. In other words, no device-level correlation is considered.
In Fig. \ref{fig:events_rate_no_space_correlation} we plot the average events rate for IDQN, MADSPG, ALOHA and TDMA protocols. We notice that TDMA has a superior performance over ALOHA. This is because TDMA guarantees collision-free time slots. However, in the case of ALOHA, in addition to redundant events, devices could also collide. On the other hand, we see that IDQN performance decreases as function of $1-\mu$ which shows that the proposed algorithm exploits efficiently the time correlation aspect of the events. For low time correlation, IDQN has similar performance to TDMA which is the preferred protocol in this regime due to its lower complexity. Finally, MADSPG outperforms all benchmarks as it can efficiently learn both stochastic and deterministic policies. Note that a deterministic policy is optimal in case of full time correlation while a uniform policy where actions are picked uniformly at random is optimal in the absence of time correlation.\par
\begin{figure}
	\centering
	\includegraphics[height= 6 cm, width= 7.5 cm]{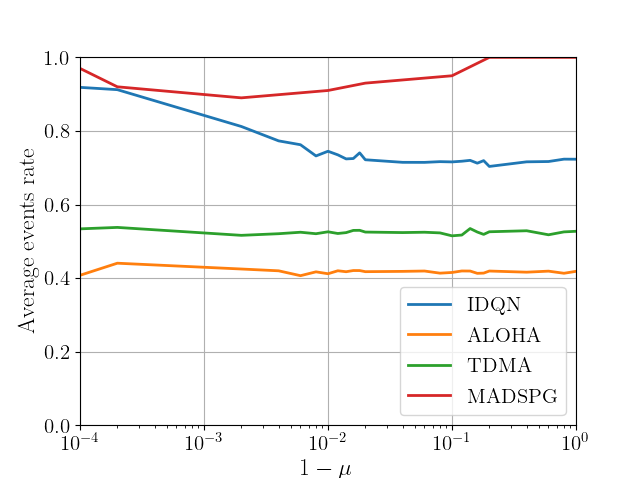}
	\caption{Average events rate for full device-level correlation scenario.}
	\label{fig:events_rate_full_space_correlation}
	\vspace{- 5mm}
\end{figure}

%\begin{figure}[t]
%	\centering
%	\includegraphics[height= 8 cm, width= 11 cm]{Figures/events_rate_one_event_active_8_devices.PNG}
%	\caption{Extreme scenario with one event being always active, $K=8$ devices, $T=3$ slots.}
%	\label{fig:events_rate_one_event_8_devices}
%\end{figure}
In order to assess the effect of device-level correlation on the system performance, we consider the set-up in Fig. \ref{fig:toy_example_full_correlation} where each device monitors all events. In Fig. \ref{fig:events_rate_full_space_correlation} we plot the average events rate for different values of $1- \mu$. We notice that in this case TDMA and ALOHA achieve lower events rate with respect to the previous case with no device-level correlation. This is because increasing device-level correlation increases the chances of redundant packets for TDMA and collisions for ALOHA. In contrast, IDQN achieves a higher performance compared to the previous case which shows that in addition to exploiting time correlation, IDQN is able to leverage device-level correlation. Similar behaviour can be observed for MADSPG which remains the best protocol with an average events rate of $92\%$. In summary, using MADSPG, each device is able to learn to transmit a suitable event by avoiding redundancy and collision with other devices transmissions by exploiting both time and device-level correlations at the cost of a higher training time due to the presence of 4 DNNs to train.
% \begin{figure}[t]
% 	\centering
% 	\includegraphics[height= 7.3 cm, width= 9.5 cm]{Figures/with_and_without_correlation_rate_episode_number.png}
% 	\caption{Average events rate under IDQN and MADSPG as function of the number of episode for no device-level and full device correlation scenarios with always active events ($q=1, p=0$).}
% 	\label{fig:events_rate_no_space_correlation}
% \end{figure}
%%%%%%%%%%%%%%%%%%%%%%%%%%%%%%%%%%%%%%%%%
\section{Conclusions}
\label{sec:conclusions}
This paper proposes a multi agent reinforcement learning algorithm for event based spectrum access. In addition to avoiding collisions and sending redundant event information, the algorithm is able to exploit efficiently time and device-level correlation of monitored events and devices respectively due to the use of a centralized critic and a stochastic policy. Numerical comparisons with state of the art solutions for spectrum access show the advantages of our algorithm. As a potential extension, we mention the derivation of reinforcement learning strategies based on the notion of correlated equilibrium \cite{correlated_equilibrium_debbah} where the correlation is based on a common external signal which would help reducing the overhead.
% \vspace{-2 mm}
\bibliographystyle{IEEEtran}
\bibliography{biblio} 
\end{document}